# Phanerozoic biological reworking of the continental carbonate rock reservoir

Craig R. Walton[1,2], Oliver Shorttle[1,3]

1 - Institute of Astronomy, University of Cambridge, Madingley Road, Cambridge, CB3OHA, UK,
2 - Department of Earth Sciences, Institute fur Geochemie und Petrologie, ETH Zurich, NW D 81.2, Clausiusstrasse 25, 8092 Zurich, Switzerland
3 - Department of Earth Sciences, University of Cambridge, Downing Street, Cambridge CB2 3EQ,UK

Understanding the co-evolution of complex life with Earth's geology is an enduring challenge. The rock record evidences remarkable correlations between changes in biology and the wider Earth system, yet cause and effect remain unclear. Here, we link the evolutionary history of eukaryotes with the rise and fall of carbonate rock fraction within continental crust – a key variable in controlling the efficiency of carbon drawdown during weathering, solid Earth degassing rates, and ultimately nutrient supply to life. We use geospatial database analyses to demonstrate a strongly non-linear growth and then collapse in Earth's continental crust carbonate reservoir. Biomineralisers reshaped Earth's surface in their image; armouring continental margins with carbonate platforms, such that the continental carbonate reservoir increased in size by 5-fold in under 100 Myr after the Cambrian Radiation of animal life. This Paleozoic carbonate revolution represents among the most dramatic crustal evolutionary events in Earth's history. The Permo-Triassic extinction event coupled to the rise of open ocean calcifiers initiated a steady decline in continental crustal carbonate content; one that still continues today, which unabated would produce Precambrian-style crustal carbonate distributions in around 500–1000 Myr. Our results demonstrate strongly non-linear crustal evolution after the rise of the complex Phanerozoic biosphere. This outcome suggests that complex life may generate unique biogeochemical trajectories on otherwise geologically similar worlds, posing a new challenge in the hunt for life beyond Earth.

Introduction

The biosphere is both influenced by and influences the geosphere. Geological processes such as mountain building and volcanism control the rate of supply of materials for weathering and ultimately nutrient release for biological consumption at Earth's surface [1]. In turn, the activity of life preferentially sequesters nutrients as well as certain mineral phases in specific environments [2], which may in the long-run have consequences for the materials that make up mountain belts and feed volcanism [3]. However, these relationships are challenging to quantify. This difficulty is in large part due to the lack of quantitative constraints on the volume and compositions of rock types being formed and destroyed at Earth's surface over time, as well as the role played by biology in shaping crustal evolution.

Recent advances in the systematisation of geological data allow for progress in this area. Large-scale geospatial, chemical, and temporal databases enable us to quantify the absolute and relative abundance of rock types in preserved continental crust and to weight the contribution of individual rock units to the average chemical composition on the basis of their respective volumes [4, 5, 6]. Critically, these tools allow us to make sense of a rock record that is a patchwork; massively varying in completeness and volume as a function of age. Here, we consider insights from a series of these growing regional-to-global databases (Macrostrat, Sedimentary Geochemistry and Paleoenvironments, i.e., SGP, and GeoROC) of crustal chemistry in concert [7, 5, 8].

The relative abundance of silicate and carbonate rock types in Earth's crust is a crucial factor underpinning the efficiency of climate-regulating and nutrient-supplying weathering processes in global biogeochemical cycles [9]. It has long been discussed that the evolutionary history of life, via carbonate biomineralisation, may have impacted the partitioning of Earth's crustal carbonate rock flux between relatively short-lived oceanic reservoirs (suffering subduction into the mantle) and comparatively long-lived storage on the continents [10, 11, 12]. Recognising such a change would be enormously important for our understanding of Earth's complex biogeochemical evolutionary history. However, no study has so far quantified the absolute and relative size of Earth's crustal carbonate reservoirs over deep time.

Here we quantify step-changes in continental crustal carbonate abundance following the advent and subsequent radiation of biomineralising animal life, which resulted in successive shifts in the locus of carbonate deposition. First, we reconstruct the evolution of the continental crustal carbonate reservoir, finding a Paleozoic upsurge and an equally dramatic Mesozoic–onward decline. Using coeval records of siliciclastic and igneous continental rock flux, we place changes in our reconstructed carbonate rock reservoir into context. We rule out solely geological interpretations of the data and instead isolate plausible geobiological mechanisms behind continental carbonate ascent and fall, highlighting a dual role for mass extinctions and the rise of novel forms of calcifying life. Finally, we explore the implications of non-linear continental crustal evolution for the search for complex life beyond Earth.

**Methods**

The Macrostrat database is organised into stratigraphic columns with a geographic footprint, each of which is associated with unique units (10). Macrostrat 'columns' – the dataset used in our analysis – are constructed from a combination of surface-exposed geology and subsurface information. Thus, Macrostrat columns typically reflect both the general geology of the surface and the subsurface down to the deepest known rock units in each region. As such, they provide some measure of crustal geology that is more inclusive than estimates that are based only on geological maps. Rock units recognised on geological maps that span multiple columns will appear as distinct units within each of those columns.

Coverage is currently limited to North America and parts of South America and New Zealand. First-order results from Macrostrat for the volume over time of preserved igneous and sedimentary rock, respectively, compare well with independent global estimates from Ronov et al. [13]. Similarly, average crustal compositions calculated using Macrostrat compare well to those calculated using explicitly global database efforts such as SGP [14, 3]. These credentials support the specific utility of Macrostrat for studying secular changes in the relative abundance of rock types in Earth's (upper) continental crust. All data are obtained with calls to the Macrostrat application program interface (API), e.g., as follows:

'https://macrostrat.org/api'.

Rock volumes were tabulated by multiplying unit thicknesses by areal extent, and dividing this total volume linearly across the reported temporal extent of the unit, i.e., time between maximum and minimum age, rather than a best estimated age with some uncertainty. This age model is currently the most self-consistent approach available in macrostrat. An obvious caveat is that, whilst sediments are laid down continuously, large volume igneous rocks may form discretely. However, our approach describes well the uncertain formation times of more ancient igneous rocks, and removes any methodological bias from the present result.

We obtain unit age-volume distributions for siliciclastic sedimentary rocks. The underlying data distribution was randomly resampled with replacement 1000 times, weighting for unit volume, to generate uncertainties. This approach to uncertainty estimation is reasonable on several counts. We are interested in our approach in calculating the total preserved volume of rock from iteratively larger spans of Earth history. As such, it is necessary to estimate the total volume of units from a given timespan, rather than the spread of data or mean. Macrostrat cannot and does not perfectly represent subsurface rock geometry, yet constraints are limited at present on the uncertainty associated with individual unit volumes. We must therefore employ an alternative means to uncertainty estimation.

Resampling with replacement in effect estimates the resulting suite of total volumes per unit time that might be obtained if the underlying data were collected several times over, i.e., in our case, if the Earth's surface were mapped and digitally represented with a database (Macrostrat) repeatedly, each time returning a slightly different result. This offers insight into how sensitive derivative results from Macrostrat are to the presence or absence of e.g., particularly volumetrically large individual units.

Using the quantitative lithological proportion metadata present in Macrostrat, we subdivided the stratigraphic thickness of units in the sedimentary rock dataset between siliciclastic and carbonate types. Using the meta-data tag information present in Macrostrat, we further subdivided the carbonate into Dunham classes. Estimates of total and tropical flooded continental area were taken from [15], extracted and imported into a python environment using the PlotDigitizer tool.

Reconstructions of carbonate flux accounting for loss of ancient crust due to erosion are obtained by upscaling the volume of ancient carbonate rock packages by the factor difference between average Phanerozoic crustal volume per Myr and the average crustal volume per Myr of the timespan in question. We use the empirical structure of the rock record in Macrostrat to define two spans of time that require distinct treatment in this way: 538.5–700 Ma, and 700 Ma and earlier.

Future evolution of Earth's continental carbonate reservoir was estimated by assuming constant preservation of carbonate rock in the same proportion as averaged over the last 15 Myr. For the status quo, we conservatively assume no loss of ancient crust in that time. We also consider a scenario where a hypothetical exhumation event on the scale of that which characterised the formation of the Great Unconformity takes place over a period of 90 Myr into the future. We consider ancient crust (i.e., that comprising the present day upper continental crust, as sampled by Macrostrat) to experience a total reduction in volume of 90%, accrued by 1% per Myr.

Results

Figures 1 and 2 plot the lithological and compositional make-up of preserved continental crust over time. Consistent with previous reports of carbonate rock volumes and the Ca contents of sedimentary rocks [10, 12], we find that the raw volume of preserved carbonate rock is much higher in the Cambrian to Permian (Paleozoic) than in the preceding Neoproterozoic (Fig. 1). However, using Macrostrat, we are able for the first time to interpret the significance of this step-change in the wider context of Earth's evolving continental crust, i.e., concurrent changes in the preserved volumes of both igneous rocks and siliciclastic sediments. Using this approach, we show that the relative volume of carbonate rock in the sum total of preserved continental crust increased sharply in the mid-to-late Cambrian period (Fig. 2a–b). This Paleozoic carbonate revolution led to a continental crustal carbonate reservoir more than 5-fold larger than that which prevailed at any point in the Precambrian (Fig. 2a).

The raw volume and relative abundance of carbonate rock declines steeply during both the End-Permian and Triassic hot-house mass extinctions (Fig. 2). In fact, the relative and absolute volume of preserved carbonate rock has never recovered to its pre-Triassic heights (Fig. 2a–b). The volume of non-carbonate sedimentary rock remained broadly stable during the Phanerozoic, with an apparent secular increase in siliciclastic rocks being offset by a decline in their derivative metasedimentary rocks (Fig. 3c). These changes manifest in a pronounced Paleozoic increase and subsequent Mesozoic–onward decrease in crustal carbonate (Fig. 2, 3a). The upsurge in sedimentary rock fraction of preserved continental crust entering the Phanerozoic is therefore driven almost entirely by changes in the continental crustal carbonate rock reservoir.

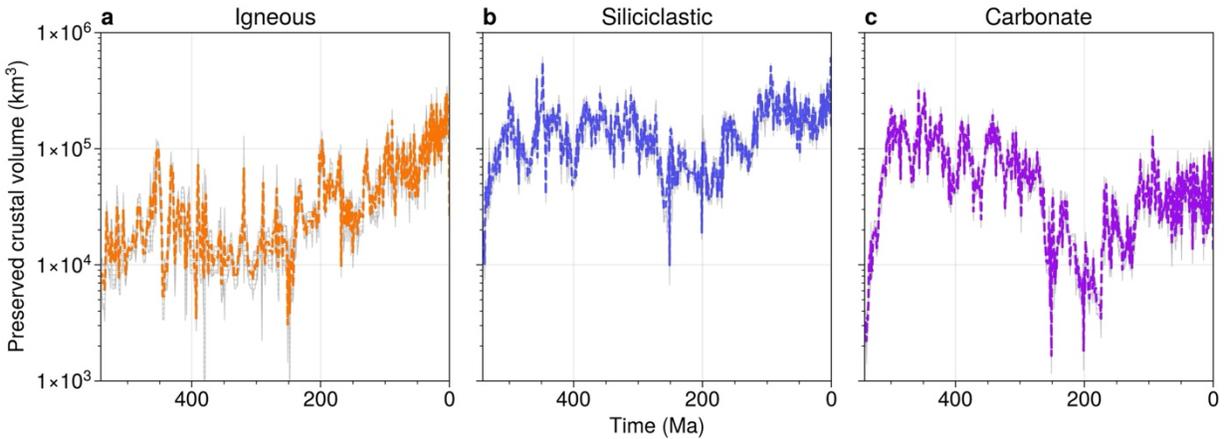

Figure 1: Secular rends in rock-type raw volume abundance and relative proportion of cumulative volume in Macrostrat database. One standard deviation errors (inherited from uncertainty in the age and volume of individual units) are shaded with hatched fill. Mean values are plotted in darker shade. The time-step is 1 Myr.

Support for the timing and magnitude of the changes in crustal carbonate that we identify is provided by the average CaO content of global fine grained sediment compositions extracted from both GeoROC [12] and the Stanford Geochemistry and Paleoenvironments (SGP) database [8] (Fig. 2b). The key shared features in carbonate rock fraction and CaO content are an initial
Cambrian upsurge, a two-peak structure in the Paleozoic, and a collapse in the Mesozoic onward

(Fig. 2b). The presence of similar geochemical signals in bulk crustal lithological reconstructions (our work) and fine-grained sediment geochemical compilations strongly supports that Macrostrat records signals that are real and global in nature, and related to secular change in carbonate production/preservation on continental crust.

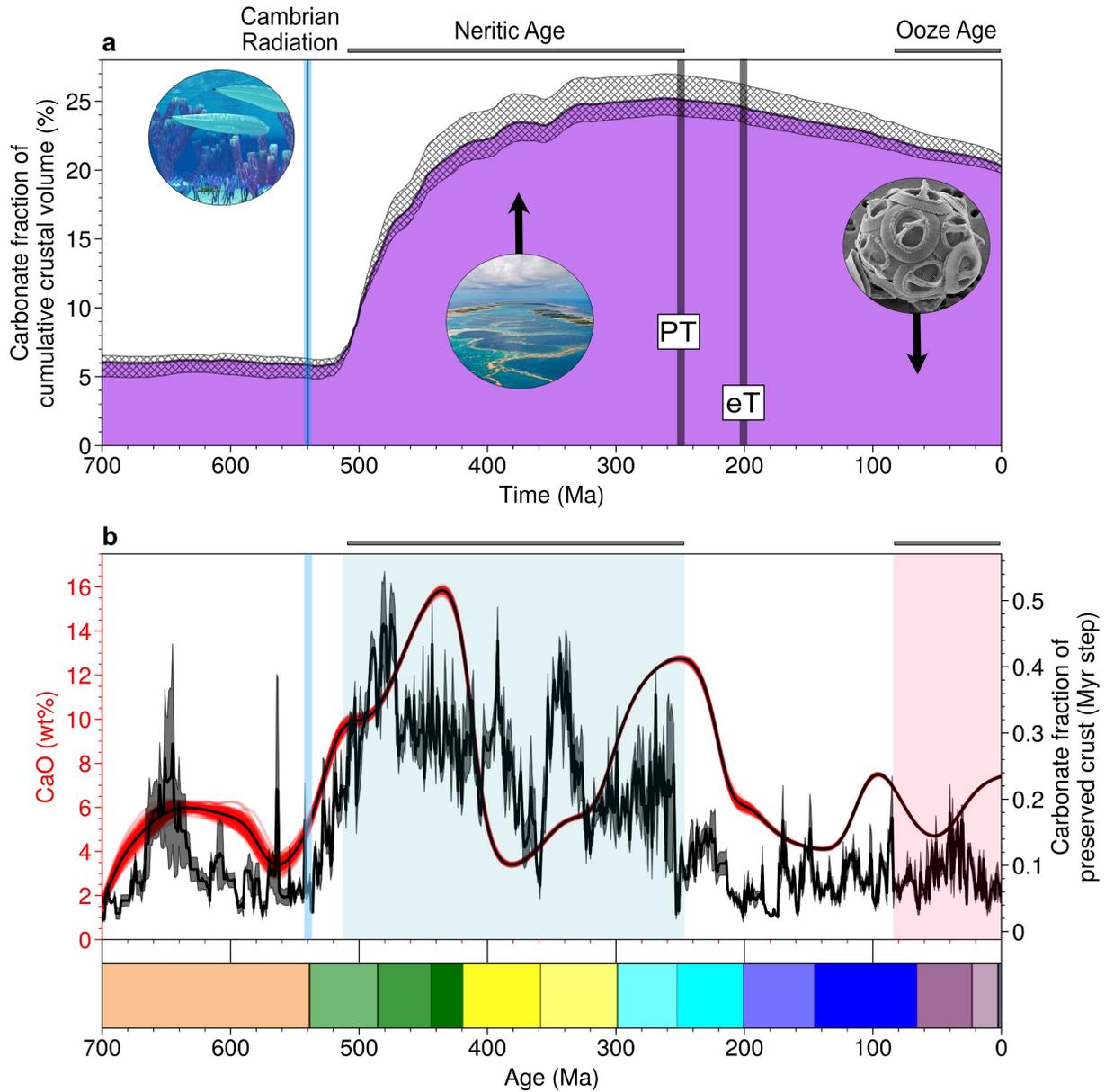

Figure 2: Evolution of the continental crustal carbonate reservoir. a) Cumulative carbonate fraction in continental crust begins to increase steeply shortly after the Cambrian radiation of animals [16] – peaks during the Carboniferous-Permian periods, and has been steadily declining ever since. PT refers to the Permo-Triassic extinction event. eT refers to the end-Triassic extinction event. b) Secular change in carbonate in coeval crustal packages preserved per Myr. Average CaO content (of 1000 uniformly resampled time series data, plotted in red) of fine-grained sediments from SGP displays the same key features as the record of crustal preserved carbonate fraction from Macrostrat and CaO records extracted from GeoROC [12].

The first order observation from Macrostrat of starkly higher relative abundance of preserved continental crust as carbonate in the Paleozoic relative to the Precambrian (Fig. 2–3) and, conversely, much lower carbonate preservation in the Mesozoic–present than in the Paleozoic requires a mechanistic explanation. We construct and test end-member hypotheses with the potential to explain the above observations. First, we consider the incompleteness of the rock record.

Apparent changes in the proportion of sedimentary rock within preserved rock packages over time may have arisen from lithology-specific preservation biases induced during erosion. In previous work, we showed that lithological proportions are stable for multi-billion-year tracts of Earth history [3]. Combined with growing evidence for tectonic rather than lithological control over rock preservation [17], we argue here that slowly accumulated bias owing to different weatherability of rock types is not responsible for the shift in preserved carbonate rock fraction at the Cambrian boundary. However, it remains possible that a major preservation bias was induced in the Proterozoic rock record at around this time due to massive erosion associated with formation of the Great Unconformity (GU). The GU is defined by a global set of unconformable contacts and between Phanerozoic and Precambrian rock units [18], indicating major crustal exhumation during the Neoproterozoic (Fig. 3c).

The missing time encoded by the GU must presumably once have been largely represented by now lost Precambrian sedimentary rocks. Based on zircon Hf-O records, an unprecedented subduction event of sedimentary material occurred coeval with the formation of the GU [19]. It may therefore be reasonable to suggest that the rock record has been somewhat biased by the preferential loss of sediment-rich Precambrian upper crustal sections during formation of the GU. However, the deeper crustal sections exhumed during formation of the GU appear to be correspondingly rich in metasedimentary rocks [3]. Indeed, the sum of metasedimentary and siliciclastic volumes represent a relatively invariant fraction of total preserved continental crust over time (Fig. 3a, 3b) [3]. Erosion has therefore biased (relative to Phanerozoic sections) only the record of low versus high grade metamorphism of siliciclastic materials, not the overall sedimentary fraction.

Moreover, a focused erosive event is not expected to have biased the ratio of siliciclastic to carbonate sediments in ancient preserved continental crust. Indeed, consistent with previous reports [20, 21], we find that Precambrian sedimentary rocks are carbonate-poor compared to Paleozoic successions (Fig. 3b). Crucially, Macrostrat results indicate that Precambrian successions contain similar carbonate rock fractions to sedimentary rocks from the Triassic onward (Fig. 3b). As such, the Cambrian upsurge in preserved carbonates is unlikely to be an artefact created during the loss of ancient sedimentary material, during the GU or otherwise. Instead, the Paleozoic Era increase in crustal sedimentary rock fraction can be attributed largely to higher rates of carbonate production and/or preservation at this time.

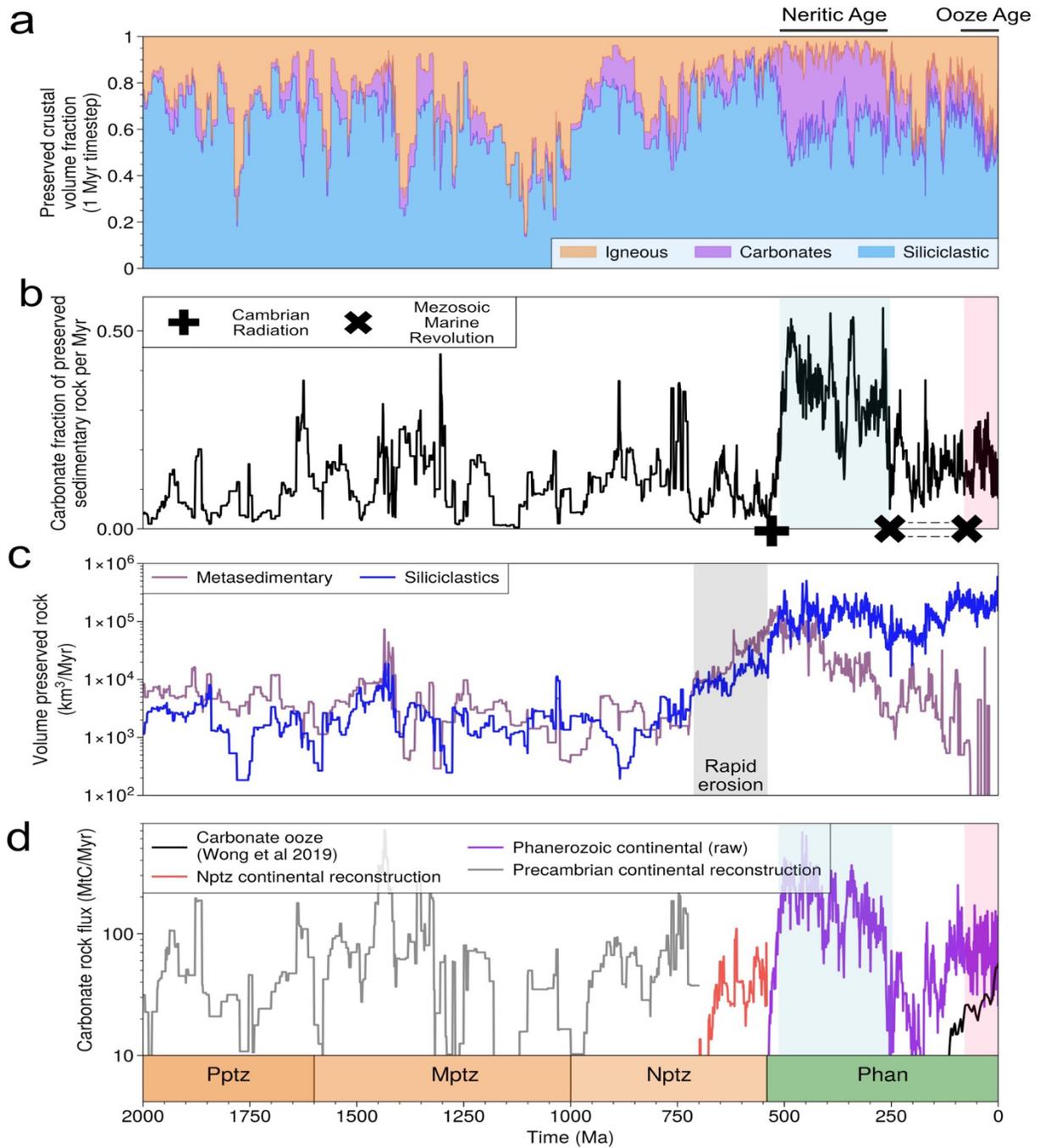

Figure 3: Earth's continental sedimentary rocks over time. a) Macrostrat volume proportions of igneous, carbonate, and siliciclastic (including metasediments) rock in preserved continental crust per Myr timestep. b) Carbonate proportion of all preserved sedimentary rocks per Myr timestep as sampled by Macrostrat. The proportion of non-carbonate sedimentary rock is broadly stable over Earth history, with the exception of the Neritic Age, which is carbonate-rich. c) The raw volume of preserved metasedimentary and siliciclastic rock sampled by Macrostrat. The fixed ratio of non-carbonate sedimentary rock (outside of Nertic Age) occurs in spite of a rise in the relative proportion of preserved rock as metasediments with increasing age, indicating that siliciclastic rocks and metasedimentary rock volumes reconcile well in a mass balance relationship due to the latter being derived from the former. d) The flux of preserved carbonate rock flux per Myr timestep, as sampled by Macrostrat. Precambrian values are corrected for erosion (see Methods). The rise of deep sea carbonate ooze since the end Cretaceous is also shown (data from Wong et al. [22]).

Consistent with our findings, the constancy of siliciclastic rock fraction over time sampled by Macrsotrat suggests that tectonic processes have operated similar to today since at least the late Archean (Fig. 3b–c). Changes in the geographic distribution of continents might instead be considered as a mechanism to change the rate of carbonate rock production, e.g., by perturbing the available flooded continental area or palaeolatitude to favour or disfavour coastal carbonate deposition. However, flooded continental area is estimated to be essentially invariant during the Cambrian period, during which a large component of our reconstructed increase in crustal carbonate fraction occurs (Fig. 4; Kocsis and Scotese [15]). In contrast, it is well documented that biology has changed substantially in that timeframe [16].

### Discussion

Geobiological perturbation of carbonate reservoirs

We propose a geobiological mechanism that links the diversification of complex life with carbonate deposition loci: the rise and fall of neritic (coastal) carbonates. Eukaryotic calcifiers are known from the Cambrian radiation of animals onwards [16], which coincides with the steep rise in carbonate rock fraction in our results (Fig. 2, 3a–b). Prior to the Triassic, calcifying life was broadly restricted to coastal benthic settings, such that their activity concentrated carbonate deposition around continental margins [16]. These materials would have been preferentially preserved on continental crust rather than subducted (Fig. 5). In contrast, Precambrian carbonates may have been largely deposited in deep sea basins [23, 21, 24, 25, 12], such that the rise of neritic biomineralisation induced a stark increase in the fraction of carbonate rock cycled into continental crust (Fig. 5).

The post-Paleozoic decline in continental carbonate rock fraction initiates coincident with the Permo-Triassic (PT) extinction, which strongly impacted calcifiers [16]. The PT boundary timeframe itself manifests in our data as a time of negligible carbonate preservation (Fig. 3a, 5). This decline is exacerbated following the end-Triassic (end-T) extinction event, another window of scarce carbonate flux (Fig. 3a). Crucially, our results show that carbonate rock preservation in continental crust never recovered from this event, with fluxes remaining lower than those in the Paleozoic from the Mesozoic-onward. The void in carbonate deposition left by the decline of neritic biomineralisers has gradually been filled by the rise of open-sea calcifying organisms, i.e., the Mesozoic Marine Revolution (Fig. 3, 5c) [22]. It is well documented that today open-sea calcifiers induce biogenic carbonate deposition in distal subduction-prone locations, creating deep sea calcareous ooze deposits (Fig. 5c) [26]. This process has been a globally important carbonate sink since the end Cretaceous [22] (Fig. 3d).

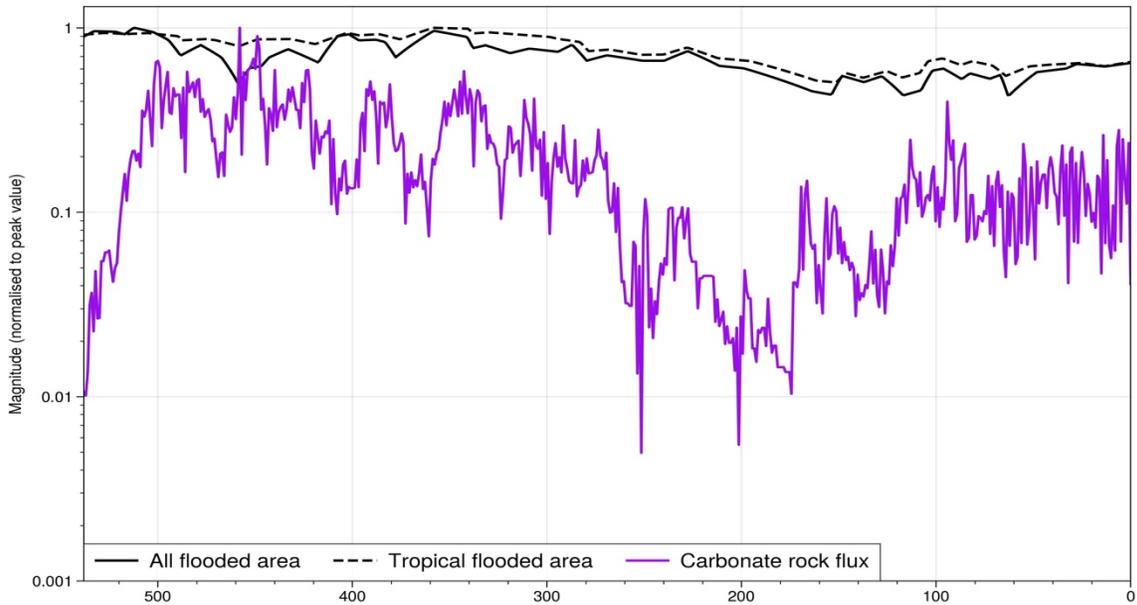

Figure 4: Variation in flooded continental area is not a likely candidate to explain observed variation in preserved carbonate rock flux. Crucially rapid and multiple order-of-magnitude changes in carbonate rock flux occur over the Cambrian, whilst flooded area is essentially invariant. Flooded area estimates are from [15].

We suggest a four part history for carbonate deposition on Earth: (1) the Precambrian was characterised by a combination of neritic (coastal) biogenic (stromatolite) carbonate and abiotic carbonate deposited in oceanic crust and upper mantle via hydrothermal alteration; (2) the radiation of animal biomineralisers shifted the locus of carbonate deposition to coastal settings for the duration of the Cambrian–Permian, such that complex neritic carbonate ecosystems reigned throughout the Paleozoic [24] – Earth's 'Neritic Age'; (3) the Mesozoic was characterised by generally low continental carbonate rock preservation; and finally (4), whilst the fraction of continental crust preserved as carbonate has remained relatively low, the rise of deep sea carbonate ooze has seen a sharp rise in oceanic crustal carbonate since the end Cretaceous [22] – Earth's 'Ooze Age'.

We test whether the estimated changes in the global flux of carbonate rock preserved on continental crust from Macrostrat are consistent with independent estimates of fluxes in Earth's carbon cycle over time (Fig. 3d, 6). The Macrostrat estimate of average mass of carbonate rock preserved per year in the Neritic Age is 135 MtC/yr versus 60 MtC/yr in the Ooze Age. These values are consistent with recent independent estimates of carbonate cycling over the last 200 Myr [22], which suggest that carbonate deposited on/in oceanic crust represented 30 MtC/yr in the Neritic Age and 60 MtC/yr in the Ooze Age.

Totalling each set of values, these results reveal rough mass balance in carbonate burial when comparing the Paleozoic (165 MtC/yr) and Cenozoic (150 MtC/yr) carbon cycles. This outcome is consistent with existing constraints for there being no clear trend in the protolith weathering intensity of sedimentary rocks throughout the Phanerozoic [14], as well as with estimates of inorganic C burial today [27]. Similarly, when scaled up by the difference in preserved average volume of preserved Phanerozoic versus Precambrian crust per Myr, average Precambrian carbonate rock flux is within error of that in the Mesozoic onward, whilst the Neritic Age still stands alone as the time of highest continental carbonate rock flux in Earth history (Fig. 3d, 6).

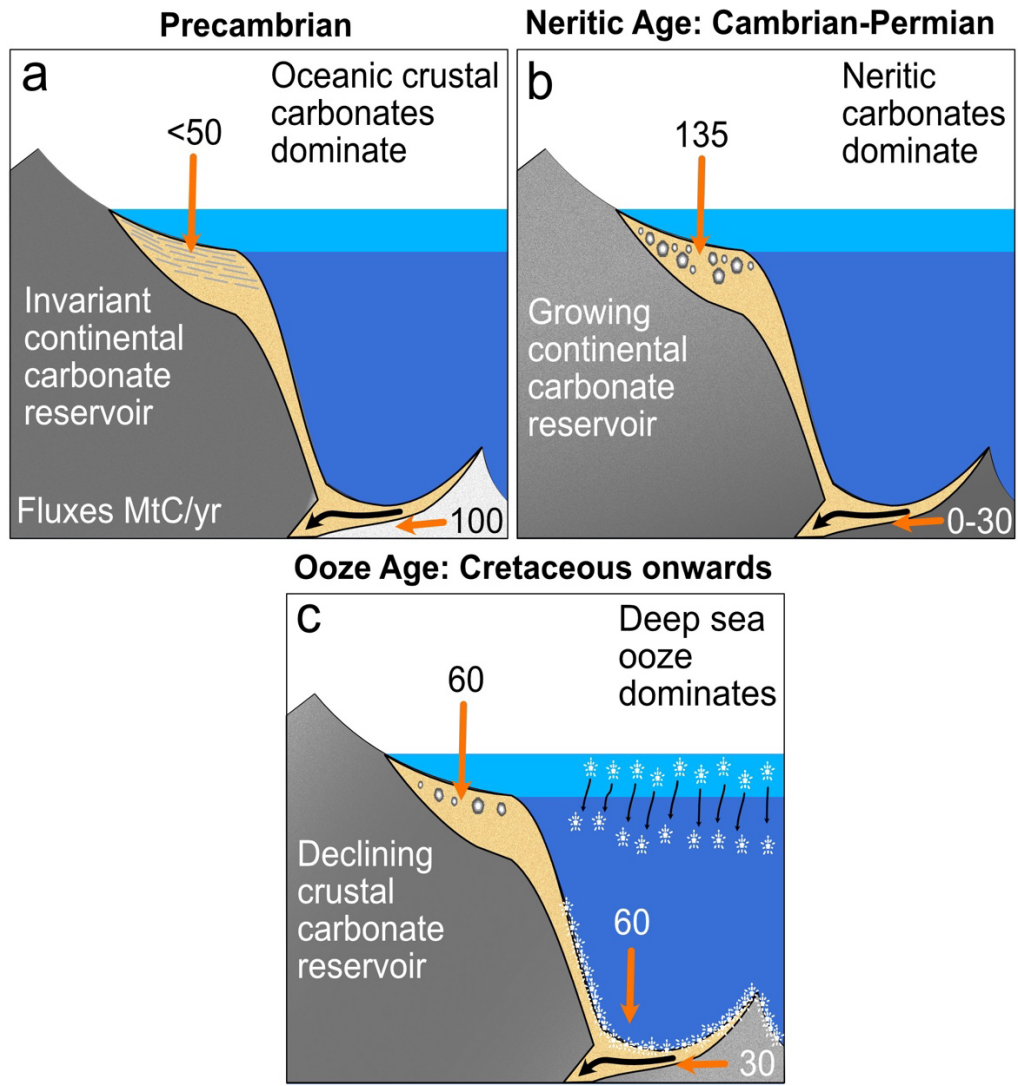

Figure 5: Non-linear crustal carbonate reservoir evolution. a) Schematic illustration of the Precambrian inorganic C cycle. Carbonate is deposited by stomatolite reefs in neritic settings and within oceanic crust. b) Neritic Age inorganic C cycle. Neritic carbonates largely or completely balance-out volcanic C degassing, with deep sea and oceanic crust carbonate burial being minor or negligible. c) The Ooze Age inorganic C cycle has been characterised by increasing C fluxes represented by open sea calcifying organisms, e.g., coccolithophores and some foraminifera. Mass balance with volcanic outgassing is achieved when considering the sum of neritic, open sea biological, and deep sea + crustal inorganic C sinks.

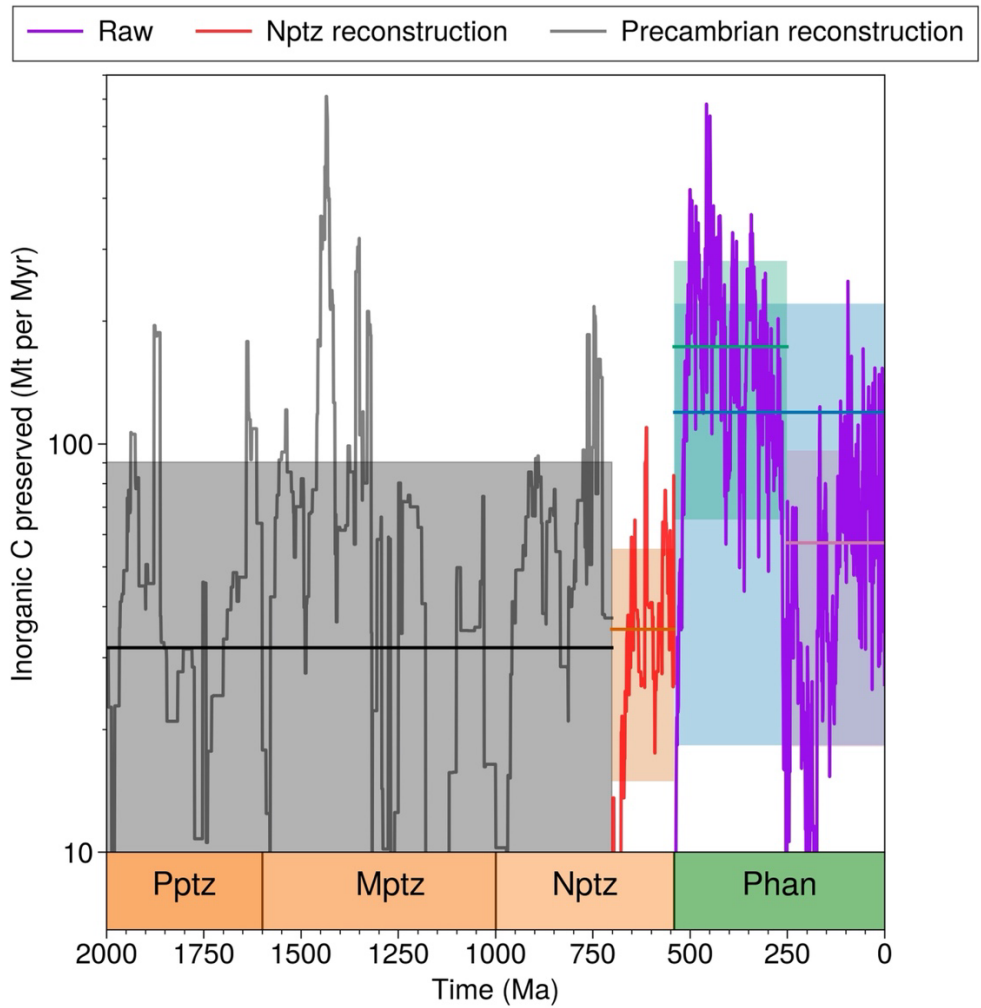

Figure 6: Precambrian reconstructed continental carbonate rock flux is within error of the Mesozoic-onward continental carbonate rock flux. The Paleozoic 'Neritic Age' stands out as the time of highest sustained continental carbonate rock flux in Earth history. Spikes in carbonate rock flux in the Precambrian may be overestimated due to the crude reconstruction approach used (see methods), which does not account for possible changes in continental volume and freeboard over time.

Our results appear to speak to the broad structure of Phanerozoic climate change. Dominant greenhouse to hothouse conditions during the Triassic-Cretaceous is consistent with net carbon outgassing following the decline of calcifying life after the PT and end-T extinction events. Importantly, the subsequent sharp Cenozoic increase in oceanic carbonate does not appear to have come at the expense of continental carbonate flux, such that the general trend of Cenozoic cooling may be mainly related to the rise of deep sea carbonate sinks during Earth's Ooze Age. Indeed, in accounting for preserved continental as well as oceanic crustal carbonate, we find that the Cenozoic has been mostly characterised by net ingassing, rather than the net outgassing previously reconstructed by Wong et al. [22]. This result supports previous claims that open sea calcifiers are driving a slow denudation of Earth's surface carbon reservoirs, via the subduction of carbonate-rich ooze [11] (Fig. 5c).

Our results suggest that carbonate sinks have been dominated by biological activity at least since the Cambrian Radiation of animal life (Fig. 3d). Importantly for Earth's carbon cycle, the rise of continental carbonate deposition in the Neritic Age increased the absolute and relative size of the continental crustal carbonate reservoir, whilst post-Paleozoic ecological dynamics have acted to reverse that trend (Fig. 2, 5). This is a crucial example of strongly non-linear continental crustal evolution, having been forced mainly by a series of biological innovations.

The rise of crustal carbonate fundamentally reorganised the Earth system, with ramifications reaching far beyond the marine realm. The ratio of carbonate to silicate rock surface availability for weathering is a critical determinant of the efficiency of regulatory feedbacks in Earth's carbon cycle [9]. However, this ratio is broadly fixed in existing biogeochemical models of Earth's evolution. Our work demonstrates that secular variation in the ratio of silicate to carbonate rock in emergent crust must be incorporated into future models, with massive and non-linear change taking place across the Phanerozoic Eon. A larger continental crustal carbonate reservoir may in principle have driven higher rates of solid Earth degassing. Continental carbonate may degas during sulfuric acid weathering, volcanic decarbonation, and orogenic metamorphism [28, 29, 30, 31].

Crustal evolution: predestined or unpredictable?

Our results suggest that animals unwittingly engineered their own environment at the vast scale of the continental crust itself. However, Earth's present carbonate-rich continental crust may also have a geobiologically-imposed expiry date. At present rates of crustal reworking, if ecological dynamics remain unchanged, crustal carbonate fraction will return to the Precambrian baseline in around 500-1000 Myr (Fig. 7a). This process would be accelerated if a widespread episode of low-temperature crustal reworking were to take place, e.g., global glaciation, widespread dynamic topography development, or an interval of elevated plate tectonic activity [18, 32, 33]. The longevity of carbonate-rich continental crust is therefore dependent on the largely unpredictable interplay of ecology, climate, and tectonics.

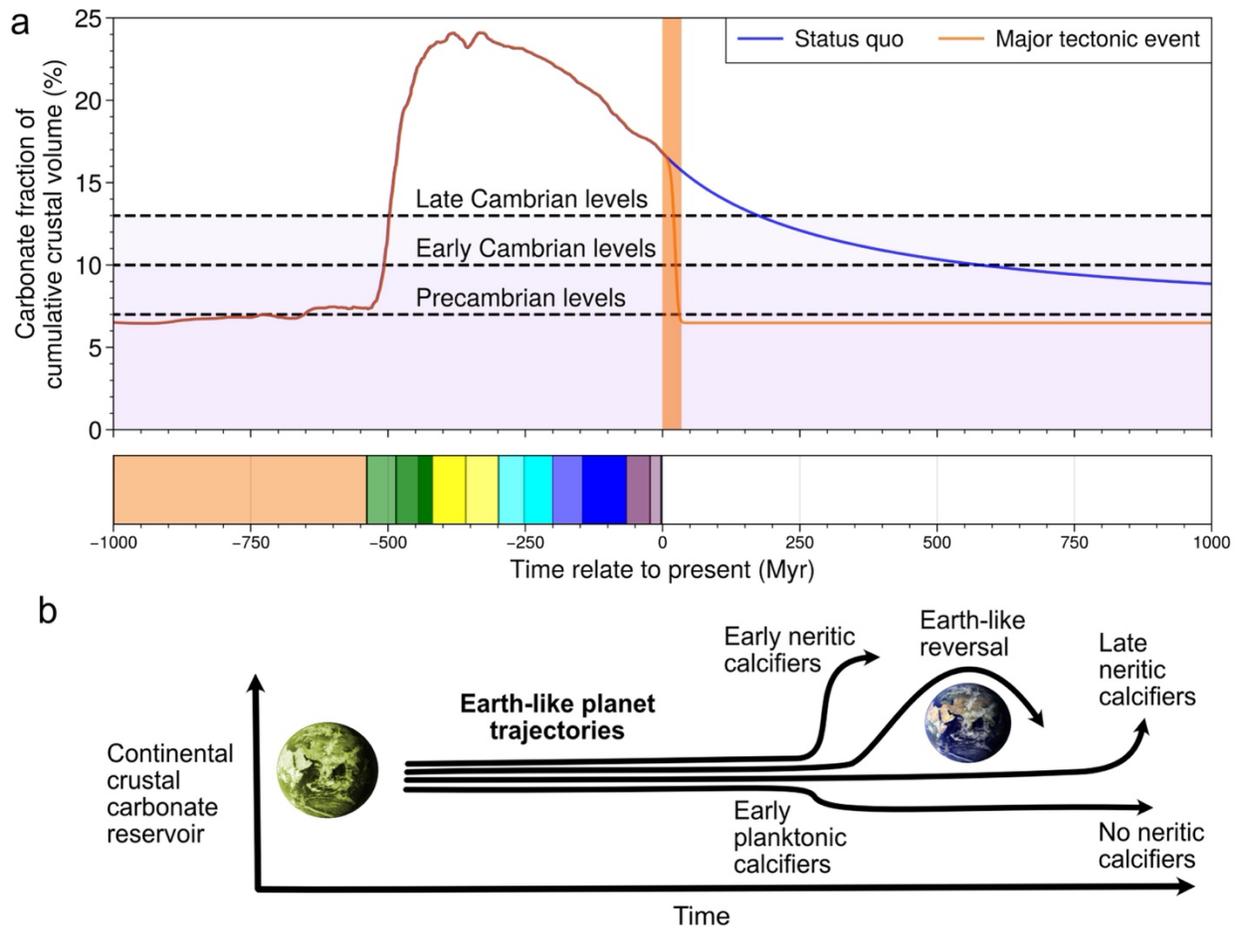

Figure 7: Possible scenarios for continental carbonate reservoir evolution a) Earth's future continental carbonate reservoir is likely to shrink if the ecological dynamics of the Ooze age are maintained. This shrinking may be relatively prolonged, given the status quo rate of Ooze age crustal reworking, or rapid, given an exhumation event of the same scale as the formation of the Great Unconformity (see methods). b) Earth-like planet crustal

evolution may be broadly predictable and linear prior to the evolution of complex life. However, complex biospheres enter into a new regime where continental crustal composition is highly sensitive to changes in the ecology of biomineralisers. In this way, random perturbations of the biosphere (e.g., asteroid impacts, large igneous province eruptions, stochastic episodes of evolutionary innovation) may translate into a non-linear and largely unpredictable path for continental crustal evolution.

Our results show that the history of life has had important ramifications for the carbon cycle and supracrustal evolution. Building on this result, future biogeochemical models must bridge two fundamentally distinct phases of Earth history: an early regime ($> 4.0$ to $0.54$ Ga), during which global planetary change occurred slowly and was determined largely by metabolic pathways, followed by the extant regime ($0.54$ Ga to present) characterised by relatively rapid and non-linear changes driven by the ecological dynamics of complex life (Fig. 7b).

The non-linearity of continental crustal evolution following the rise of complex life is important in the search for life in the universe. Crucially, the long-term evolution of planets hosting complex biospheres may be challenging to predict from local astrophysical and geochemical context alone. Ultimately, whilst the initial evolution of a group of geological and biologically Earth-like planets may be similar, we may expect strong divergence depending on the local evolutionary history of planktonic vs neritic calcifying organisms (Fig. 7d). Predicting and identifying the distribution of complex life in the cosmos will therefore require us to reckon with a vast parameter space of non-linear ecological niche discovery. This challenge is imposing but perhaps not insurmountable, with Earth's Phanerozoic Eon providing us with a detailed fiducial case of non-deterministic planetary evolution.

### Conclusions

Earth's emergent continental crust represents the integrated product of igneous, sedimentary, metamorphic, and biological activity over billions of years. The carbonate fraction of the crust is both a record of past changes in volcanism, Earth's climate, and biological niche discovery, as well as an active participant in forcing global biogeochemical change. By assessing geo-spatial, -temporal, and -chemical data, we identify an upsurge in the carbonate fraction of the preserved rock record at the start of the Phanerozoic Eon. By normalising to trends in the preserved volume of igneous and siliciclastic sedimentary rocks, we show that this upsurge cannot be explained by preservation bias. Instead, we provide evidence in support of a key role for the bioengineering of the crust, with animal life acting to enhance carbonate preservation on continental margins. This trend reversed following the widespread extinction of many neritic calcifying organisms alongside the rise of open sea calcifiers, which shifted the locus of carbonate burial back to subduction-prone deep marine settings. We speculate that otherwise Earth-like worlds may see entirely different evolutionary trajectories in continental carbonate content subject to the whims of biological innovation versus extinction.

Acknowledgements: C.R.W. acknowledges NERC and UKRI for support through a NERC DTP studentship, grant number NE/L002507/1; financial support from the Cambridge Leverhulme center for Life in the Universe; funding support from Trinity College (Cambridge) in the form of a Junior Research Fellowship; and funding support from ETH Zurich and the NOMIS foundation in the form of a research fellowship. We thank Nick Butterfield, Nicholas Tosca, Tony Prave, Sasha Turchyn, Ed Tipper, and Eva Stueeken for keen insights during the early development of this work. Martin Walton is thanked for inspiring the direction of this work at a crucial early stage.

Data Availability All data needed to evaluate the conclusions of this manuscript are available in the National Geoscience Data Centre (NGDC) permanent repository. Data are accessible via the following link: *(data currently being processed)*